\documentclass[11pt]{article}

\usepackage[]{ACL2023}

\usepackage{times}
\usepackage{latexsym}

\usepackage[T1]{fontenc}

\usepackage[utf8]{inputenc}

\usepackage{microtype}
\usepackage{pdflscape}
\usepackage{rotating}
\usepackage{amssymb}
\usepackage{inconsolata}

\AtBeginDocument{%
  }

\usepackage{amsmath,amsfonts}
\usepackage{algorithmic}
\usepackage{graphicx}
\usepackage{textcomp}
\usepackage{xcolor}
\usepackage{longtable}
\usepackage{makecell}
\usepackage{subfigure}
\usepackage{xspace}
\usepackage{float}
\usepackage{tipa}

\usepackage{arydshln}
\usepackage{enumitem}
\usepackage{multirow}
\usepackage{tcolorbox}
\usepackage{colortbl}
\usepackage{booktabs}

\usepackage[T1]{fontenc}
\usepackage{tabularx}
\usepackage{threeparttable}
\usepackage{relsize}

\begin{document}

\title{LogicAsker: Evaluating and Improving the Logical Reasoning Ability of Large Language Models}


\author{
    {\bf Yuxuan Wan}$^{1}$\thanks{\ \ Both authors contributed equally to this research.} ,
    {\bf Wenxuan Wang}$^{1}$\footnotemark[1] ,
    {\bf Yiliu Yang}$^{1}$, 
    {\bf Youliang Yuan}$^{2}$, \\
    {\bf Jen-tse Huang}$^{1}$,
    {\bf Pinjia He}$^{2}$,
    {\bf Wenxiang Jiao}$^{3}$\thanks{\ \ Wenxiang Jiao is the corresponding author.},
    {\bf Michael R. Lyu}$^{1}$ \\
    $^1$The Chinese University of Hong Kong, Hong Kong, China \\
    $^2$The Chinese University of Hong Kong, Shenzhen, China \\
    $^3$Tencent AI Lab, China \\
    \texttt{\{yxwan9, wxwang, jthuang, lyu\}@cse.cuhk.edu.hk}, \texttt{yyiliu@link.cuhk.edu.hk},\\
    \texttt{youliangyuan@link.cuhk.edu.cn}, \texttt{hepinjia@cuhk.edu.cn}, \texttt{wenxiangjiaonju@gmail.com} \\\\
}

\maketitle

\begin{abstract}
We introduce LogicAsker, a novel approach for evaluating and enhancing the logical reasoning capabilities of large language models (LLMs) such as ChatGPT and GPT-4. Despite LLMs' prowess in tasks like writing assistance, code generation, and machine translation, assessing their ability to reason has been challenging. Traditional evaluations often prioritize accuracy on downstream tasks over direct assessments of reasoning processes. LogicAsker addresses this gap by employing a set of atomic reasoning skills grounded in propositional and predicate logic to systematically examine and improve the reasoning prowess of LLMs. Our methodology reveals significant gaps in LLMs' learning of logical rules, with identified reasoning failures ranging from 29\% to 90\% across different models. Moreover, we leverage these findings to construct targeted demonstration examples and fine-tune data, notably enhancing logical reasoning in models like GPT-4o by up to 5\%. To our knowledge, this is the first effort to utilize test case outcomes to effectively refine LLMs' formal reasoning capabilities. We make our code, data, and results publicly available\footnote{\url{https://github.com/yxwan123/LogicAsker}} to facilitate further research and replication of our findings.

\end{abstract}
\newcounter{partno}
\setcounter{partno}{1}

\section{Introduction}
\label{sec-introduction}

\begin{figure*}
\vspace{-12pt}
    \centering
    \includegraphics[width=0.99\textwidth]{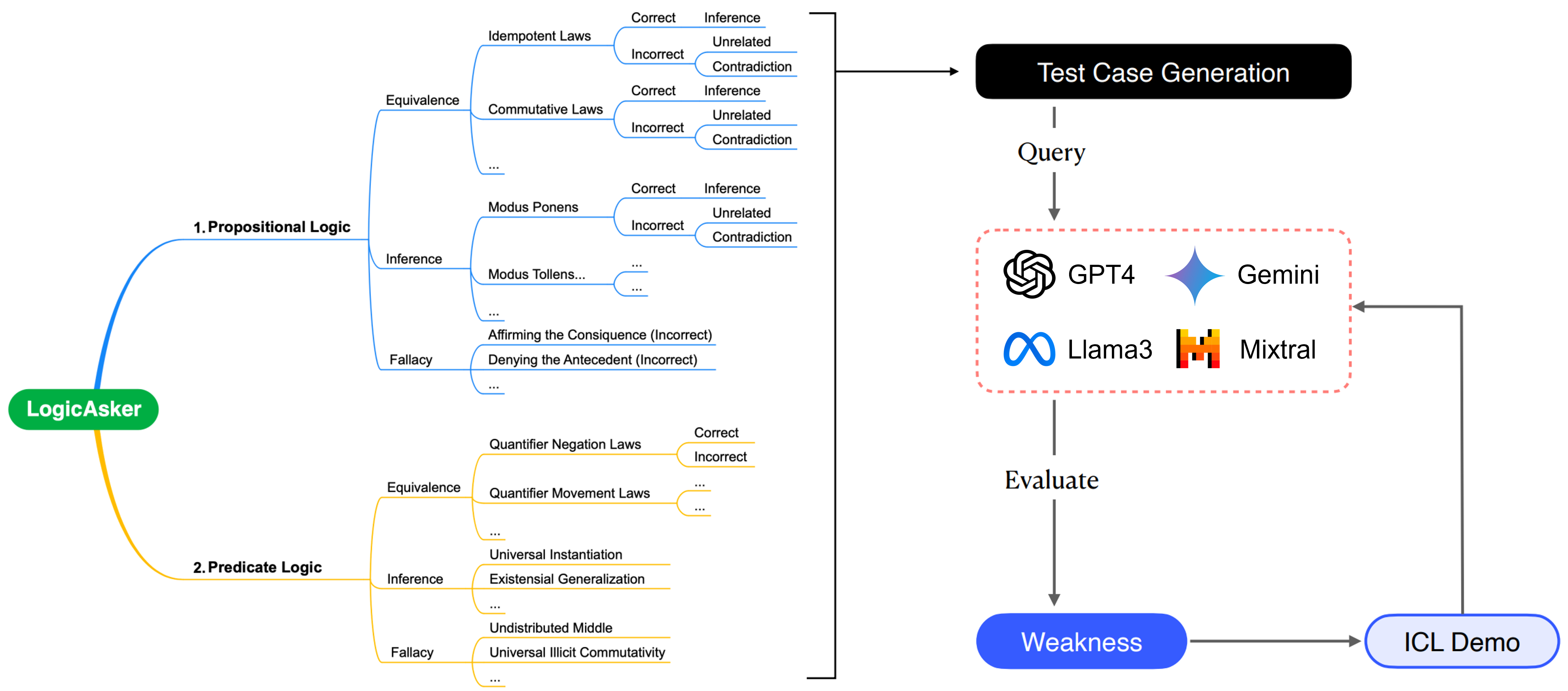}
    \caption{Overview of the LogicAsker framework.}
    \label{fig:overview}
    \vspace{-12pt}
\end{figure*}

Large language models (LLMs), such as OpenAI's GPT series have significantly impacted natural language processing, excelling in a variety of tasks including text generation, machine translation, and code generation~\cite{Gao2022ComparingSA,Gao2023ConstructingEI,Jiao2023IsCA}. 

Reasoning, defined as the cognitive process of using logic to draw conclusions from given facts~\cite{Wei2022ChainOT, Wei2022EmergentAO}, is crucial for complex interactions that go beyond text generation. Accurately assessing this ability in LLMs is essential, yet challenging, as models may correctly perform tasks merely relying on shortcuts such as pattern recognition without truly engaging in logical reasoning~\cite{Huang2022TowardsRI, Huang2023LargeLM,Liu2023EvaluatingTL}. Consider the following inference example: \texttt{Either it is raining, or Tom will play football; if it rains, then the floor will be wet; the floor is dry; therefore, Tom will play football.} We may encounter the following challenges: 1) It's unclear if a correct LLM response is due to reasoning or simple heuristics like word correlations (e.g., ``dry floor'' is more likely to correlate with ``playing football''). 2) If an LLM fails, pinpointing the specific breakdown in reasoning is difficult (i.e., inferring not raining from the floor being dry or inferring playing football from not raining). 3) Current systems lack comprehensive test cases that encompass various formal reasoning types beyond implication, such as logical equivalence (e.g., A and B are true; therefore, B and A are true. 4) Evaluating an LLM’s reasoning on such cases offers limited insight into enhancing its reasoning capabilities.

To better handle these challenges, a well-performing testing framework should be able to define a set of skills that a) directly correspond to the reasoning process, b) cannot be further divided, c) cover all formal logical reasoning scenarios, and d) can identify LLMs' weaknesses and facilitate improving LLMs' performance. Property a) ensures that the task cannot be accomplished by other approaches, such as inferring from the correlations of words, and the evaluation result directly reflects the model's reasoning ability. Property b) and c) ensure that the set of skills is fundamental and comprehensive, which can provide helpful insights to accomplish Property d).

We introduce LogicAsker, an automatic framework designed to evaluate and enhance LLMs' formal reasoning skills using Minimum Functionality Tests (MFTs)~\cite{Ribeiro2020BeyondAB}, akin to software engineering's unit tests, which utilize straightforward examples to assess specific behaviors. These tests help identify when models rely on shortcuts rather than genuinely mastering a skill~\cite{Ribeiro2020BeyondAB}. Specifically, LogicAsker builds a set of atomic skills from foundational principles of propositional and predicate logic, two fundamental systems used to formalize reasoning procedures~\cite{Partee1990MathematicalMI}, together with common logical fallacies~\cite{Hurley2020ACI}. Based on the skill set, LogicAsker generates reasoning questions by translating standard logic expressions into natural language, assesses LLMs’ accuracy per skill, pinpoints weaknesses, and creates in-context-learning~\cite{Brown2020LanguageMA} examples and fine-tuning data to bolster reasoning abilities. In addition, for each skill, LogicAsker uses diverse vocabulary to frame various natural language queries, computing average performance to minimize biases from word correlations.

Table~\ref{tab:compare} demonstrates that LogicAsker complements existing frameworks by providing a comprehensive evaluation scope and utilizing outcomes to enhance LLMs' reasoning capabilities, while other datasets often face data leakage and are scope-limited. LogicAsker serves as an extensive diagnostic tool for LLMs' formal reasoning, significantly exceeding the coverage of comparable tools and enabling detailed assessments across diverse reasoning rules such as inferences, quantifiers, and fallacies. Scaling up the scope presents significant challenges due to the complexity of designing algorithms capable of processing various logical rules and translating them into natural language. Despite these complexities, LogicAsker uniquely integrates all formal logical rules and common fallacies, facilitating robust testing and refinement of reasoning capabilities.

We evaluated LogicAsker's performance through extensive testing on six state-of-the-art (SOTA) LLMs~\cite{chatbot_arena_leaderboard}, including four closed-source LLMs (GPT-4o, GPT-4, ChatGPT, and Gemini-1.5) and two open-source LLMs (Llama3 and Mixtral). Our findings reveal that LogicAsker's test cases effectively pinpoint logical reasoning failures across these models, with error rates (i.e., $1 - \text{accuracy}$) between 29\% and 90\%. These test cases also facilitate the creation of in-context learning examples and fine-tuning data, thereby enhancing logical reasoning capabilities. For instance, applying LogicAsker's cases to GPT-4o improved its reasoning accuracy from 92\% to 97\%. All resources are released for reproduction and further research\footnote{\url{https://github.com/yxwan123/LogicAsker}}. 

We summarize the main contributions of this work as follows: 
\begin{itemize}[leftmargin=*]
    \item  We are the first work that formally defines a comprehensive set of 34 atomic and 208 extended skills necessary for LLMs to execute formal reasoning based on propositional and predicate logic.
    \item We develop LogicAsker, a fully automatic tool that utilizes atomic skills to generate test cases to assess and enhance LLMs' reasoning abilities, marking a first in utilizing test results to directly improve LLM performance.
    \item We conduct a thorough empirical evaluation of the logical reasoning abilities of six SOTA LLMs. We demonstrate that the test results by LogicAsker can be used to effectively evaluate and improve the performance of LLMs.
\end{itemize}

\begin{table*}
    \centering
    \caption{Comparison with previous works.}
    \label{tab:compare}
    \small
    \begin{tabular}{p{4.8cm} p{1.2cm} p{0.8cm} p{0.8cm} p{1.2cm} p{1.3cm} p{0.8cm} p{0.8cm} p{1cm}}
    \hline
    & Fully Automatic & Atomic Skills & Formal Rules & Include Fallacies & Identify Weakness & Improve LLMs & LLMs$^*$ Tested & Example Testbed \\
    \hline
    CLUTRR~\cite{Sinha2019CLUTRRAD} & $\times$ & $\times$ & $\times$ & $\times$ & $\checkmark$ & $\times$ & - & BERT \\
    LogiQA~\cite{Liu2020LogiQAAC} & $\times$ & $\times$ & $\times$ & $\times$ & $\times$ & $\times$ & - & BERT \\
    RECLOR~\cite{Yu2020ReClorAR} & $\times$ & $\times$ & $\times$ & $\times$ & $\checkmark$ & $\times$ & 2 & GPT2 \\
    Soft Reasoner~\cite{Clark2020TransformersAS} & $\checkmark$ & $\times$ & 1 & $\times$ & $\checkmark$ & $\times$ & - & RoBERTa \\
    LogicNLI~\cite{Tian2021DiagnosingTF} & $\times$ & $\times$ & 7 & $\times$ & $\checkmark$ & $\times$ & - & BERT \\
    FOLIO~\cite{Han2022FOLIONL} & $\times$ & $\times$ & $\times$ & $\times$ & $\times$ & $\times$ & 4 & GPT3 \\
    LogicInference~\cite{Ontan2022LogicInferenceAN} & $\checkmark$ & $\times$ & 19 & $\times$ & $\times$ & $\times$ & - & T5 \\
    ProntoQA-OOD~\cite{Saparov2023TestingTG} & $\checkmark$ & $\times$ & 6 & $\times$ & $\checkmark$ & $\times$ & 4 & GPT3.5 \\
    \hline
    \textbf{LogicAsker (Ours)} & $\checkmark$ & $\checkmark$ & 34 & $\checkmark$ & $\checkmark$ & $\checkmark$ & 6 & GPT4 \\
    \hline
    \end{tabular}
    \raggedright\footnotesize{* We consider language models with more than 1 billion parameters as LLMs.}
\end{table*}

\section{Preliminaries}
\label{sec-backgound}

\subsection{Formal Analysis of Reasoning Abilities}
\label{sec:formal-analysis}
``Reasoning'' can be characterized into formal reasoning and informal reasoning. The former is a systematic and logical process that follows a set of rules and principles, and the reasoning within these systems will provide valid results as long as one follows the defined rules (e.g., all A are B, all B are C; therefore, all A are C). The latter is a less structured approach that relies on intuition, experience, and common sense to draw conclusions and solve problems ~\cite{Huang2022TowardsRI, Bronkhorst2020LogicalRI} (e.g., Hong Kong residents have a high life expectancy; this is probably because they have healthy living habits). Generally, formal reasoning is more structured and reliable, whereas informal reasoning is more adaptable and open-ended but may be less reliable. In this paper, we focus on the formal reasoning process to systematically analyze LLMs' reasoning abilities. 

To formalize reasoning procedures, two fundamental systems are usually adopted, namely, propositional logic and predicate logic. The former one deals with propositions or statements that can be either true or false, and utilizes logical operators including $\land$ (and), $\lor$ (or), $\neg$ (not), $\rightarrow$ (inference), and $\leftrightarrow$ (bidirectional) to connect these statements. The latter one, in contrast, extends propositional logic to deal with more complex statements that involve variables, quantifiers, and predicates.
Both propositional logic and predicate logic contain various rules for the reasoning process. 
These rules can be categorized into equivalence rules and inference rules. Equivalent rules summarize the basic expressions that are equivalent in terms of truth value (e.g., $\neg(P \land Q)$ $\Leftrightarrow$ $(\neg P) \lor (\neg Q)$). Inference rules summarize the basic valid inference rules (e.g., from the premises:  $A \rightarrow B$, and $A$, we can infer $B$ ). 

We refer to~\cite{Partee1990MathematicalMI} for a more detailed explanation. Table~\ref{tab:prop-equiv}-\ref{tab:inference-law} in Appendix \ref{appendix:rules} list common inference rules in predicate logic and propositional logic. Besides inference rules, formal logic systems can also express common logical fallacies, i.e., arguments that may sound convincing but are based on faulty logic and are, therefore, invalid. We list the common logical fallacies in Table~\ref{tab:common-fallacies}.

\subsection{Minimum Functionality Test}
In this paper, we adopted the concept of Minimum Functionality Tests (MFTs), introduced in~\cite{Ribeiro2020BeyondAB}, to evaluate the reasoning ability of LLMs. MFTs are analogous to unit tests in software engineering, where a collection of simple examples is used to check a specific behavior within a capability. These tests involve creating small and focused datasets that are particularly effective in detecting whether models resort to shortcuts to handle complex inputs, rather than truly mastering the capability.

To apply MFTs in evaluating the reasoning ability of LLMs, we treated each formal logical rule as an independent task and generated abundant test cases for each task. Each test case was designed to trigger logical failures in the LLMs, allowing us to assess the strengths and weaknesses of LLMs in the logical reasoning process, and providing a solid foundation for further analysis and improvement.

\section{LogicAsker}
\label{sec-method}

\begin{figure*}
    \centering

    \includegraphics[width=0.98\textwidth]{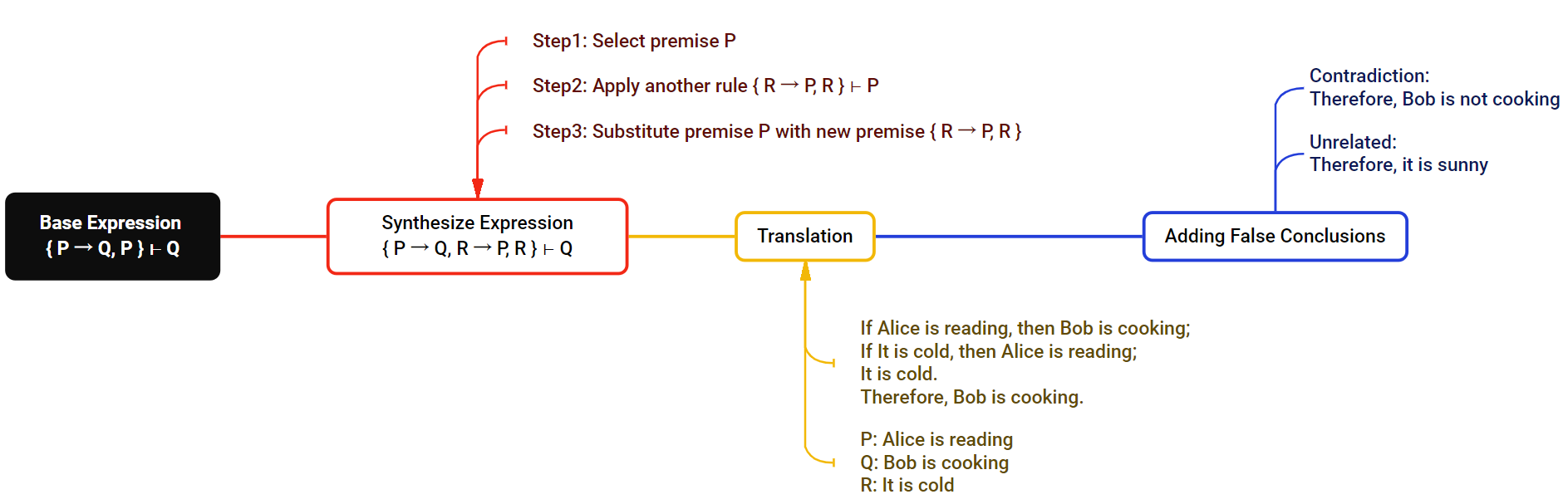}
    \caption{Test case generation procedure.}
    \label{fig:generation_flow}
    \vspace{-12pt}
\end{figure*}

In this section, we introduce the design and implementation of LogicAsker, a novel tool to trigger logical reasoning failures in large language models. Figure~\ref{fig:overview} overviews the workflow of LogicAsker, which consists of three main modules: test case generation, weakness identification and in-context learning (ICL) demonstration. In particular, the test case generation module utilizes atomic skills defined on the two formal logic systems and an inference synthesis approach to generate questions as test cases. Then, the generated cases are fed into the LLMs to reveal weaknesses and provide insights into the 
LLMs by the weakness identification process. Finally, LogicAsker utilizes these insights to construct ICL demonstrations to improve the reasoning abilities of the LLMs.

\subsection{Reasoning Skills}
\textbf{Atomic skills.} As described in Section~\ref{sec:formal-analysis}, propositional and predicate logic are two fundamental systems that formalize the reasoning process. The inference rules and equivalence laws in these two systems are atomic and can cover all correct reasoning scenarios; therefore, we define these 34 rules as the set of atomic skills an LLM should possess to perform formal reasoning.

\textbf{Extended skills.} Predicate logic extends propositional logic to deal with more complex statements that involve variables, quantifiers, and predicates. In this regard, besides the unique equivalence and inference laws in predicate logic, we add quantifiers and variables to every rule in propositional logic to form the predicate version of the laws. Using this approach, we expand the set of 34 atomic skills into a set of 208 extended skills. In Appendix \ref{appendix:extended-rules}, we provide some concrete examples of these extended rules.

\subsection{Test Case Generation}
\label{subsec:testcase_gen}
To generate logical questions, LogicAsker first adopts a rule-based method to generate logical expressions systematically based on reasoning skills and then translates the logical expressions into natural language.  Figure~\ref{fig:generation_flow} provides an overview of the procedure.

\textbf{Logic expression generation.} To better control the process of logic expression generation, we first define the length of an inference problem by the number of syllogisms it involves.
We use the inference rules described in Section~\ref{sec:formal-analysis} to generate inference expressions with length one.
When a longer inference ($> 1$) is specified, we start with a base expression $E_0 := P_1 \land P_2 \rightarrow C_1 $ with length one and expand the inference chain.
Specifically, we substitute the premises (either or both) of the first inference with the conclusion of some other syllogism and append the premises of those syllogisms into the list of all premises.
For example, we can find another syllogism $E_1 := P_3 \land P_4 \rightarrow P_2$ with $P_2$ as the conclusion and then obtain a new expression $E_{new} := P_1 \land P_3 \land P_4 \rightarrow C_1$ with the inference length of two. 
We can obtain inference expressions of any length by recursively expanding the inference chain as above.
During the generation process, one can specify the desired rules and length to allow complete control over expected test cases.

In addition to the correct inference expression created above, we generate three kinds of false inference expressions: contradiction, unrelated, and fallacy.
A contradiction is generated by negating the conclusion of a correct inference expression and an unrelated is generated by replacing the conclusion of a valid inference expression with an irrelevant statement.
For example, for $E_0 := P_1 \land P_2 \rightarrow C_1$, a contradiction is $E_c := P_1 \land P_2 \rightarrow \neg C_1$, an unrelated can be $E_u := P_1 \land P_2 \rightarrow U_1$.
We create a fallacy by directly using the fallacy rules listed in Section~\ref{sec:formal-analysis} for an inference length of one. 
For a fallacy with a more extended length, we select a fallacy rule as the base expression and expand the inference chain using correct rules, ensuring the expression's incorrectness.



\textbf{Natural language translation.}
Partially inspired by~\cite{Ontan2022LogicInferenceAN}, translating a clause into natural language involves a series of patterns that depend on the structure of the clause. Simple propositions are transformed into one of the template patterns, such as ``subject verb-action'', ``subject predicate'', or ``impersonal-action'' with a predefined set of subjects, verbs, predicates, and impersonal actions that can be chosen randomly without repetition. For predicate clauses that involve constants or variables, we employ templates ``subject verb-action'', ``subject predicate'' to translate them. Furthermore, each clause can be rendered in various modes, such as the present, past, or negated forms. Additionally, connectives like "or," "and," "implies," and "if and only if" also adhere to their designated patterns. For quantified clauses, we adopt patterns like "for all $x$, $X$", "there is at least one $x$ for which $X$", and "some $X$s are $Y$,". To facilitate the generation process, we curate extensive lists of potential subjects, including common names in English, and compile plausible predicates, actions, and impersonal actions. Compared to Ontanon et al. ~\cite{Ontan2022LogicInferenceAN}, our method provides a more natural and less ambiguous translation. We provide a detailed illustration of the translation process in Appendix \ref{appendix:translation}.

\subsection{Weakness Identification}
\label{sec:metrics}
To measure the reasoning abilities of the LLMs, we calculate the accuracy of LLMs' answers $\text{Acc} = \frac{N_{\rm correct}}{N_{\rm total}}.$ Where ${N_{\rm total}}$ denotes the total number of responses, and $N_{\rm correct}$ denotes the number of responses that are correct. In particular, since all generated queries are formulated as yes-or-no questions, LogicAsker adopts an automatic approach that searches for pre-defined keywords (e.g., "yes" and "no") in sentences to identify correct answers. 

To reveal the weaknesses of LLMs, we generate $n$ test cases for each leaf node in the rule tree depicted in Figure 1. Then, we calculated the response accuracy of an LLM of each leaf node. Based on the result, we can identify the weaknesses of LLMs by listing the leaf nodes that receive the lowest accuracy. In addition, by grouping the accuracy by different attributes in the rule tree, we can gain insights into the strengths and weaknesses of LLMs on these attributes (e.g., performance on predicate logic vs. propositional logic). 

\subsection{Improving LLMs}
\textbf{In-context learning (ICL)} is a paradigm that enables LLMs to learn tasks with examples in the form of demonstrations~\cite{Brown2020LanguageMA}. It leverages task instructions and a few demonstration examples to convey the task semantics, which are then combined with query questions to create inputs for the language model to make predictions. ICL has demonstrated impressive performance in various natural language processing and code intelligence. However, the performance of ICL is known to rely on high-quality demonstrations~\cite{Gao2023WhatMG} strongly. To fully unleash the potential of ICL, LogicAsker utilizes the weak skills of each LLM to construct both correct and incorrect examples with expected answers and explanations as demonstrations to facilitate the reasoning of LLMs. The
generation process follows a similar approach to the test
case generation described in § 3.2, with the difference being that we append a brief explanation and the correct answer at the end of each case. We show an instance of the demonstration example in Appendix \ref{appendix:prompting}. 

\textbf{Fine-tuning} is another widely used technique to enhance model performance on specific tasks~\cite{Moslem2023FinetuningLL, Wei2021FinetunedLM}. This process involves taking a pre-trained model and further training it on a smaller, task-specific dataset. The rationale behind fine-tuning is to leverage the learned features and knowledge of the pre-trained model, adapting it to particular nuances and characteristics of a targeted domain or task. In this paper, we directly utilize the data generated by LogicAsker to fine-tune LLMs to improve their reasoning ability. 

\section{Experiments}
\label{sec-experiment} 

\begin{table}[t]
\centering
\caption{Conversational LLMs used in the evaluation.}

\begin{tabular}{l c}
\toprule
\bf Name  &\bf Rank\\
\midrule
GPT-4o~\cite{openai_gpt4o}    & 1  \\
GPT-4~\cite{openai_gpt4} & 4  \\
ChatGPT~\cite{openai_chatgpt} & 37 \\
Gemini-1.5~\cite{google_gemini} & 10 \\
\hdashline
Llama3-70b~\cite{meta_llama3} & 12\\
Mixtral-8x7b~\cite{mistral_mixtral} & 42  \\
\bottomrule
\end{tabular}
\label{tab:chatbot}
\vspace{-10pt}
\end{table}

\subsection{Experimental Setup}

We apply LogicAsker to test six state-of-the-art (SOTA) LLMs, including four close-source and two open-source models. Table~\ref{tab:chatbot} lists brief information on these systems.
All of them are ranked within the top~50 in the LMSYS Chatbot Arena Leaderboard~\footnote{\url{https://huggingface.co/spaces/lmsys/chatbot-arena-leaderboard}} according to the assessment results in June 2024. We leave details of how we access the model, the parameters used, and the prompt we used in Appendix \ref{appendix:access-LLMs}.

We conduct two iterations of experiments for a comprehensive assessment. In the first iteration, we follow the setting in \S~\ref{sec:metrics} and set $n=25$, resulting in 5,200 cases. Statistics of the sampled data are described in Appendix~\ref{appendix:stat}. The second iteration is based on the first one, which focuses on the identified weaknesses of each LLM, i.e., the ten leaf nodes in Figure~\ref{fig:overview} with the lowest accuracy.
We generated 25 additional test cases for each weakness. These 250 test cases comprise our ``weakness dataset, '' which will be utilized for further evaluation in \S~\ref{sec-improve}. 

\subsection{Effectiveness of LogicAsker}
\label{sec:effective}

\begin{figure}[t!]
    \centering
    \includegraphics[width=.95\columnwidth,]{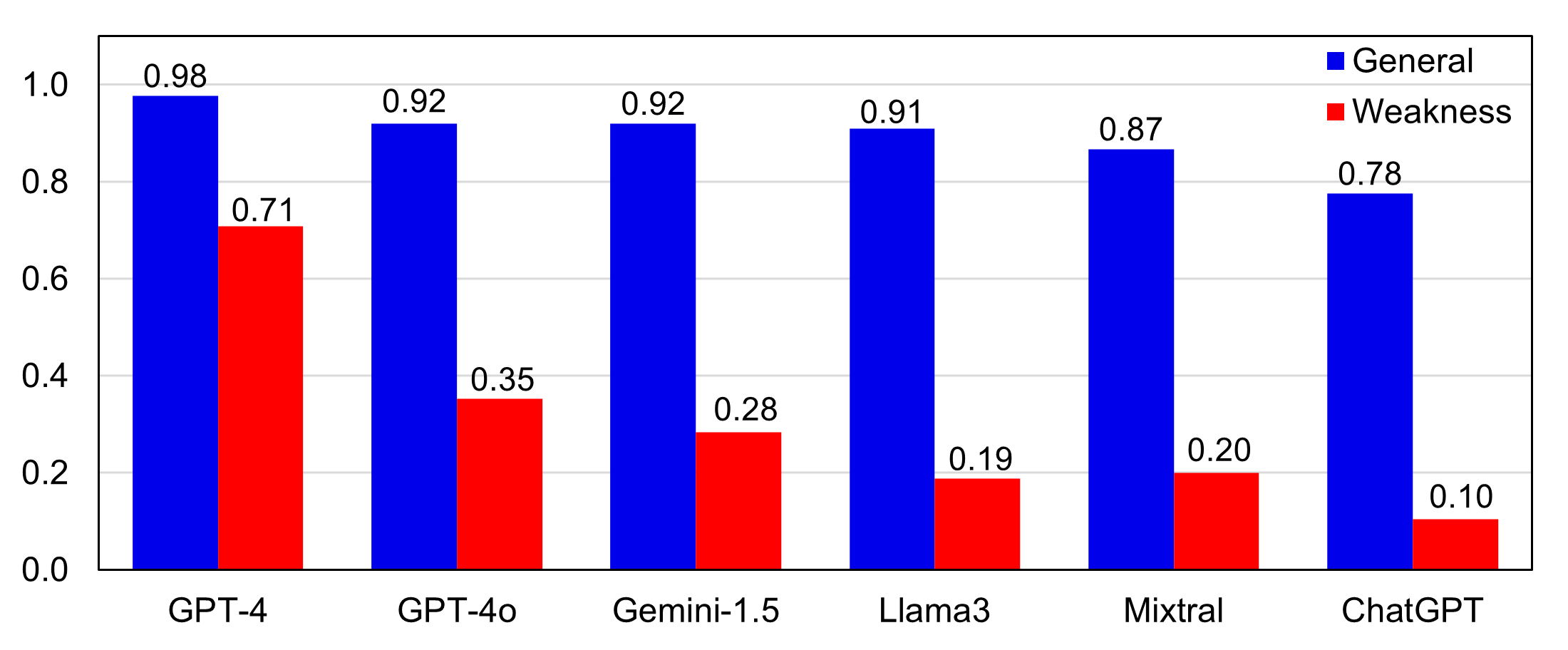}
    \vspace{-6pt}
    \caption{Overall accuracy.}
    \label{fig:overall}
\end{figure}

We demonstrate the effectiveness of LogicAsker through the overall performance of LLMs on the test cases. The overall performance of LLMs in the first and second iteration is shown in Figure~\ref{fig:overall}.
The result reveals that our framework can effectively expose logical failures in the first iteration, with LLM's accuracy ranging from 78\%-98\%. When focusing on the weak skills of LLMs in the second iteration, we further reduce the accuracy to  10\%-71\% for the LLMs.
What's surprising is that most of these LLMs show accuracy even lower than random guesses (i.e., 50\% here) when confronted with logical questions involving specific logical rules. This contradicts their remarkable performance in various LLM benchmarks, for example, achieving top~50 ranks on the LLM Arena Leaderboard.
It suggests that existing benchmark datasets are not comprehensive enough to assess the generalization ability of LLMs in reasoning.


\begin{figure}[t!]
    \centering
\includegraphics[width=.95\columnwidth]{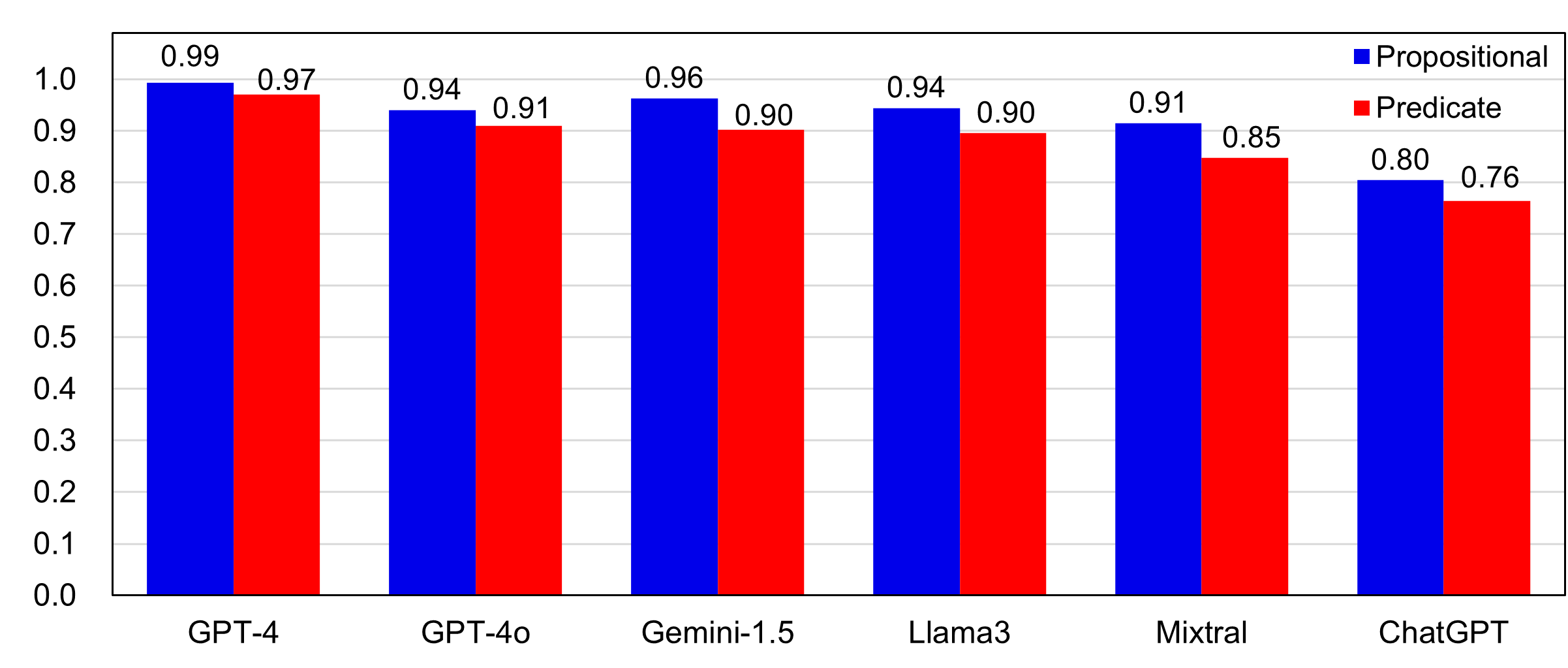}
    \vspace{-6pt}
\caption{Propositional and predicate logic accuracy.}
    \label{fig:diff_logic}
\end{figure}

\begin{figure}
    \centering

    \includegraphics[width=.95\columnwidth]{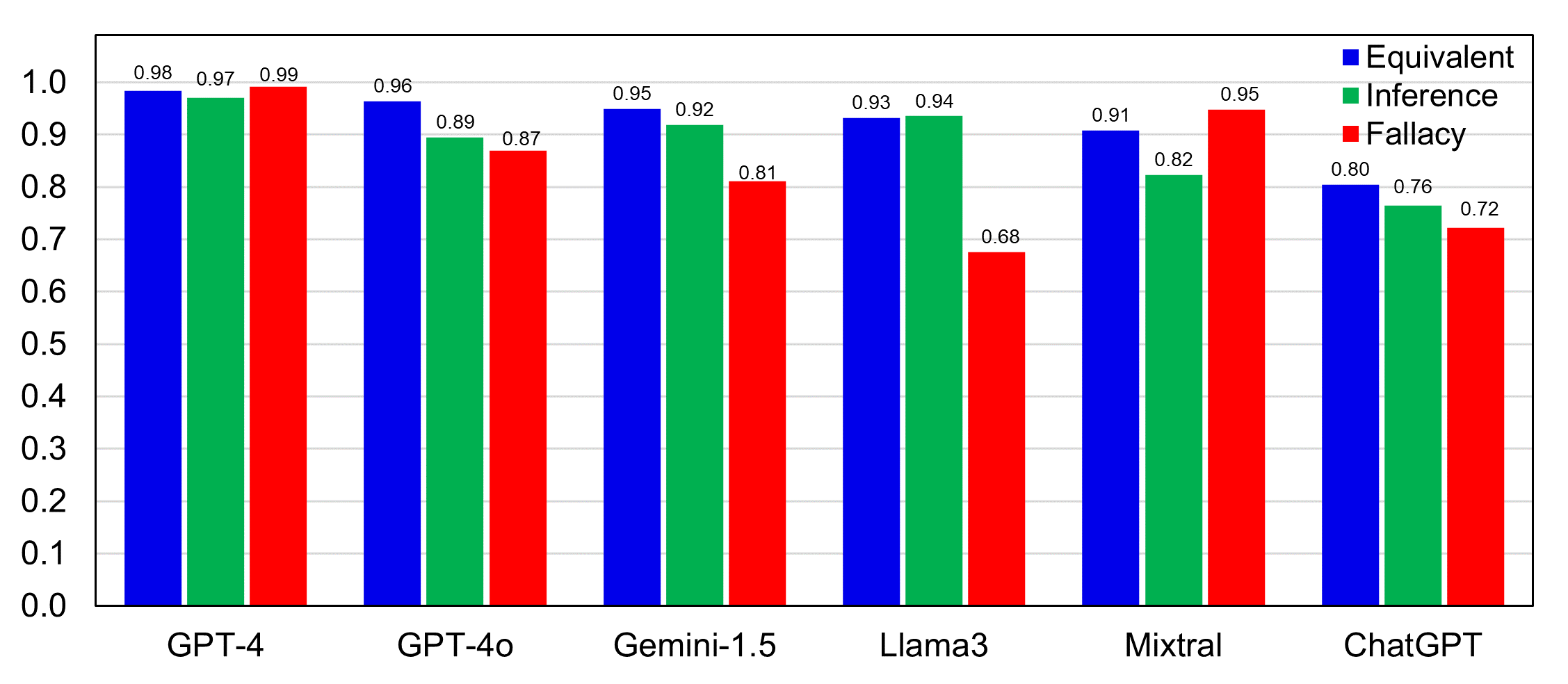}
    \vspace{-6pt}
    \caption{Accuracy of different rule categories.}
    
    \label{fig:diff_category}
\end{figure}

\subsection{Insights into Reasoning Abilities} 
We conducted a comprehensive analysis to gain insights from the failures exposed by LogicAsker, obtaining three key observations from the evaluation: 

\textbf{Most LLMs are better at easier logical skills.} 
We compared the performance of LLMs on propositional logic and predicate logic, the former of which is simper in form while the latter involves more complex quantifier manipulations.
Figure~\ref{fig:diff_logic} illustrates the difference between the accuracy obtained for the two logic systems. Notably, we observed that most LLMs are better at propositional logic, implying their limited ability in complex reasoning scenarios.

\textbf{Most LLMs are weak in recognizing logical fallacies.} Figure~\ref{fig:diff_category} presents the accuracy of LLMs under different skill categories. Interestingly, we discovered that among three types of skills, recognizing fallacies has the lowest accuracy for most LLMs, with GPT-4 and Mixtral-8x7B being the exceptions.
It suggests that current LLMs are over-confident even in fallacies, which may be learned from the mistakes in pretraining data.

\begin{table*}[ht]
\centering
\small
\caption{Weakness of GPT-4}
\label{tab:gpt4-case}
\begin{tabularx}{\textwidth}{@{}lcXr@{}}
\toprule
\textbf{Rule} & \textbf{Type} & \textbf{Example} & \textbf{Accuracy} \\ \midrule
Existential resolution & Incorrect & For all v, v will not play squash or v will go running. There is at least one v for which v will play squash or v will play tennis. Therefore, there is at least one v for which v will go running. & 0.60 \\
Universal resolution & Correct & For all x, x will climb a mountain or x is a police officer. For all x, x will not climb a mountain or x is rich. Therefore, for all x, x is a police officer or x is rich. & 0.60 \\
Law of quantifier movement & Correct & For all x, if Joseph sleeps, then x is a janitor. Therefore, if Joseph sleeps, then for all x, x is a janitor. & 0.68 \\ \bottomrule
\end{tabularx}
\vspace{-10pt}
\end{table*}

\textbf{Case study: GPT-4 did not learn all logic rules well.} To provide a direct impression of what skills LLMs cannot perform well, we list three atomic rules in which GPT-4 has the lowest accuracy in Table~\ref{tab:gpt4-case}. While GPT-4 has an average accuracy of 98\% over all skills, it only achieves 60\% - 68\% accuracy on these skills, indicating that it cannot perform these atomic skills smoothly. 

We also discovered that longer inference chains are more challenging for LLMs, the details are provided in Appendix~\ref{appendix:lengths}. These insights provide a valuable understanding of the strengths and weaknesses of each LLM when handling logical questions, allowing us to uncover specific areas that require improvement and potential avenues for enhancing overall performance. We provide a full breakdown list of the LLMs' performance on various skills in Appendix~\ref{appendix:breakdown-list}.

\begin{table}[t]
\centering
\caption{Human evaluation results on the quality of test cases.}

\label{tab:annotation}
\small
\begin{tabular}{lcccc}
\toprule
\textbf{Invalid Cases} & \textbf{a} & \textbf{b} & \textbf{c} & \textbf{Total} \\
\midrule
Count & 10 & 8 & 0 & 18 \\
Percentage & 1.92\% & 1.54\% & 0.00\% & 3.46\% \\
\bottomrule
\vspace{-10pt}
\end{tabular}

\end{table}

\subsection{The Quality of Test Cases}

Since the test cases are automatically generated, we conduct a human evaluation to measure the quality of the generated test cases by LogicAsker. To achieve this, we randomly sampled 10\% (520) of the test cases generated during the first iteration of the experiment in \ref{sec:effective} and conduct manual inspection. Two annotators with bachelor's degrees were recruited to answer the questions manually. Each test case was annotated as either valid or invalid based on the following three questions: \textbf{a)} Is the question grammatically correct? \textbf{b)} Is the question understandable and has only one interpretation? \textbf{c)} Can the target answer be derived from the question? A test case is considered valid only when both annotator's answer to the above questions are positive. The results of the annotation are presented in Table~\ref{tab:annotation}. This result is statistically sufficient to prove that the probability of LogicAsker generating understandable and solvable logical questions is larger than or equal to 0.94 (with p-value 0.05), indicating that \textbf{the test cases generated by LogicAsker are highly reliable and valid}.


\subsection{LogicAsker to Improve Reasoning}
\label{sec-improve}
In this section, we explore the potential of LogicAsker in further improving the reasoning ability of LLMs through in-context learning (ICL) and fine-tuning.

We first employ LogicAsker to generate ICL demonstrations tailored to address the weaknesses dataset uncovered in the experiments in \S~\ref{sec:effective}.
For each inference problem, we generated ICL demonstrations that provide both the expected answer and an explanation as described in \S~\ref{sec-method}.
We evaluate the effectiveness of the ICL demonstrations generated by LogicAsker by comparing the following prompting strategies:
a) Zero-Shot: We provide only task instructions without any ICL demonstrations. b) Zero-Shot Chain-of-Thouhgt (CoT): We use the instruction "Please think step-by-step"~\cite{Kojima2022LargeLM} to elicit the zero-shot reasoning ability of the LLMs. c) Random ICL Demonstrations: In addition to the task instruction, we also include four ICL demonstrations selected randomly from the available rules with balanced answer labels, i.e., two correct and two incorrect. d) Weakness ICL Demonstration: Instead of random demonstrations, we include four ICL demonstrations using the weakness rules identified in \S~\ref{sec:effective} with balanced answer labels. 

We perform testing with the 5.2k sampled data on all models and list the result in Table~\ref{tab:icl}.
In general, the weakness ICL demonstrations are more effective than those random ICL demonstrations, and both ICL methods bring more performance gain than CoT, \textbf{indicating that the test cases generated by LogicAsker can improve reasoning}. 

\begin{table}[t]
    \centering
    \caption{Performance of ICL demonstrations by LogicAsker (\%).}
    \small
    \label{tab:icl}
    \begin{tabular}{@{}lcccc@{}}
\toprule
\textbf{Model}      & \textbf{Zero-Shot} & \textbf{CoT} & \textbf{ICL}    & \textbf{ICL(Weak)} \\ \midrule
GPT-4      & 97.75    & 96.60        & 97.98 & 99.48   \\
GPT-4o     & 91.92    & 92.94        & 95.77 & 97.23   \\
Gemini     & 92.06    & 93.62        & 96.13 & 96.67   \\
Llama-3    & 91.02    & 94.54        & 94.83 & 93.35   \\
Mixtral    & 86.77    & 86.23        & 76.40 & 82.02   \\

ChatGPT    & 77.62    & 78.19        & 82.90 & 81.04   \\\midrule
Average      & 89.52    & 90.35        & 90.67 & \textbf{91.63}   \\ \bottomrule
\end{tabular}
    \vspace{-12pt}
\end{table}

To further demonstrate the effectiveness of LogicAsker, we fine-tune ChatGPT on 5.2k separately generated data on all skills and 2.8k separately generated data on weaknesses of ChatGPT, respectively. We test the two fine-tuned model on both LogicAsker and another dataset, LogiQA, a challenging dataset for machine reading comprehension with logical reasoning~\cite{Liu2020LogiQAAC}. We use the "test" split of LogiQA which contains 651 test data. The results are presented in Table~\ref{tab:chatgpt_performance}. We can observe that models fine-tuned on LogicAsker can effectively enhance the models' reasoning ability on both datasets, suggesting the generalizability of LogicAsker. \textbf{These findings demonstrate the effectiveness of LogicAsker in improving the reasoning ability of LLMs}.

\begin{table}[ht]
\centering
\caption{ChatGPT performance on LogiQA and LogicAsker after fine-tuning (\%).}
\label{tab:chatgpt_performance}
\vspace{-5pt}
\small
\begin{tabular}{@{}lccc@{}}
\toprule
 & \textbf{Vanilla} & \textbf{FT (All)} & \textbf{FT (Weak)}\\ \midrule
 LogicAsker & 77.62 & \textbf{99.50} & 97.83 \\
  LogiQA & 40.55 & 41.01 & \textbf{41.78} \\\bottomrule
\end{tabular}
\vspace{-5pt}
\end{table}

\section{Discussion}

A surprising result from our experiments is the lower performance of CoT prompting in some LLMs, such as GPT-4, despite the widespread belief in CoT's ability to enhance logical reasoning.

One possible explanation lies in the fundamental difference between how CoT operates and the structure of LogicAsker’s test cases. CoT functions by breaking down complex reasoning tasks into intermediate, manageable steps, guiding the model toward a final conclusion through a series of natural language reasoning steps. However, LogicAsker tests focus on assessing LLMs' mastery of the most basic, indivisible reasoning skills. These fundamental reasoning skills are the building blocks for more complex reasoning processes like CoT. Since these basic skills cannot be further subdivided into intermediate steps, the benefits of CoT are limited when applied to LogicAsker's simpler test cases. In contrast, improving an LLM's understanding of these core skills can enhance its performance on more complex tasks where CoT is applicable, as demonstrated by ChatGPT's improved results on the LogiQA dataset (see Table 6).

Further analysis of failure cases reveals another intriguing observation. In certain CoT-prompted scenarios, GPT-4 tends to rely on common sense rather than logical deduction. Given that LogicAsker's data is synthetic and not aligned with common-sense reasoning, this tendency leads to performance degradation. This is illustrated by the following example, where CoT prompts GPT-4 to incorrectly infer weather conditions based on premises, while a zero-shot approach provides a more accurate deduction by sticking to strict logical interpretation:

\begin{quote}
\small
\textbf{Q:} Consider the following premises: If Linda were sad, then it is sunny, and if it is sunny, then Linda is sad. Can we infer the following from them? Answer yes or no: It is not overcast.

\textbf{Ans (Zero-shot):} From these premises, we can infer that \textit{Linda is sad} and \textit{it is sunny} are equivalent; ... This means that Linda is sad if and only if it is sunny. However, without additional information about whether Linda is actually sad or not, we cannot definitively conclude the state of the weather (overcast or not). Therefore, based on the information given and without knowing whether Linda is sad, the answer is \texttt{no}.

\textbf{Ans (CoT):} ... The statement \textit{“It is sunny”} directly contradicts the possibility of it being overcast at the same time, assuming a typical understanding of weather where sunny conditions and overcast conditions are mutually exclusive. ... Therefore, if either is true, it must be sunny, and thus it cannot be overcast. ... Since the premises lead us to conclude that it is sunny, we can infer that it is not overcast. Final answer: \texttt{yes}.
\end{quote}

This observation underscores the effectiveness of LogicAsker in revealing situations where models default to heuristics or memorized knowledge rather than true logical reasoning. While CoT has been shown to improve logical reasoning in general tasks, LogicAsker's framework exposes when models fail to genuinely reason and instead fall back on familiar or remembered patterns. This insight suggests that strengthening LLMs' mastery of basic reasoning skills is a necessary foundation for improving performance on tasks that benefit from CoT strategies.

\section{Related Work}
\label{sec-related}
Significant advancements in NLP reasoning have been achieved through methods such as Chain-of-Thoughts (CoT) prompting~\cite{Wei2022ChainOT}, which enables models to generate reasoning steps with minimal training. Enhancing this, the Program of Thoughts (PoT) prompting~\cite{Chen2022ProgramOT} leverages external interpreters like Python to handle complex mathematical problems. Further augmenting reasoning validity, the Logic Agent framework transforms LLMs into logic agents that can dynamically apply propositional logic rules to convert natural language inputs into structured logic forms~\cite{Liu2024LogicAE}.

Recent studies have focused on evaluating the reasoning capabilities of Large Language Models (LLMs) by measuring their performance across various reasoning tasks. These include arithmetic \cite{Cobbe2021TrainingVT, Hendrycks2021MeasuringMP, Amini2019MathQATI, Patel2021AreNM, Miao2020ADC, Ling2017ProgramIB, Roy2016SolvingGA}, commonsense \cite{Talmor2019CommonsenseQAAQ, Geva2021DidAU, Clark2018ThinkYH}, symbolic \cite{Wei2022ChainOT}, and table reasoning \cite{Nan2021FeTaQAFT}, as well as understanding words, dates, and causal relationships \cite{Srivastava2022BeyondTI}, and generalization \cite{Lake2017GeneralizationWS, Anil2022ExploringLG}. Despite these efforts, it remains uncertain whether LLMs truly reason or rely on simple heuristics, since most assessments focus only on accuracy and do not thoroughly evaluate the reasoning processes.

Efforts to develop metrics for more formal reasoning analysis in LLMs include creating datasets with first-order logic problems \cite{Han2022FOLIONL}, generating test cases using a single predicate inference rule \cite{Saparov2022LanguageMA}, building instruction-tuning dataset designed for CoT reasoning with GPT-4~\cite{Liu2023LogiCoTLC}, and employing propositional logic with randomized methods~\cite{Ontan2022LogicInferenceAN}. These methods, however, often lack generalizability or focus on limited deduction rules. Saparov et al. \cite{Saparov2023TestingTG} introduced a comprehensive approach by using all deduction rules in propositional logic to assess LLMs' deductive reasoning across complex proofs. Our research expands further, incorporating all rules and equivalent laws in both propositional and predicate logic, aiming to enhance understanding of each rule's impact on LLM performance and using these insights for improvement.

 
\section{Conclusion}
In this paper, we present LogicAsker, an automated tool designed to comprehensively evaluate and improve the formal reasoning abilities of LLMs under a set of atomic skills. 

Our research demonstrated the efficacy of LogicAsker in identifying logical reasoning failures in a diverse set of widely deployed LLMs, we achieved a substantial error detection rate in revealing reasoning flaws in these models, ranging from 29\% to 90\%. Additionally, we utilized the test cases from LogicAsker to design in-context learning demonstrations, which effectively enhance the logical reasoning capabilities of LLMs, e.g., improving from 92\% to 97\% for GPT-4o.

By providing insights into the strengths and weaknesses of LLMs in reasoning, we are able to improve the reliability and trustworthiness of these models. The release of all the code and data aims to facilitate replication and encourage further research in this crucial area. 

\section*{Limitations}

This paper identifies two primary limitations that highlight areas for future research:
\begin{itemize}[leftmargin=*]
    \item Although our ICL (In-Context Learning) method significantly enhances the logical reasoning capabilities of large language models (LLMs), there remains a performance gap compared to human-level reasoning. Further refinements and innovations in model training and architecture may be necessary to bridge this gap.
    
    \item Our method is currently applicable only to LLMs that possess robust in-context learning capabilities. LLMs lacking this feature may not benefit from our approach. Future studies could explore fine-tuning methods to extend the applicability of our improvements across a broader spectrum of LLMs, potentially enhancing models with weaker or no inherent in-context learning abilities.
\end{itemize}

\subsubsection*{Acknowledgments}

The paper is supported by the Research Grants Council of the Hong Kong Special Administrative Region, China (No. CUHK 14206921 of the General Research Fund).

\bibliographystyle{acl_natbib}
\bibliography{reference}

\appendix

\label{sec:appendix}

\section{Logical Rules and Fallacies}
\label{appendix:rules}
We list all the logic equivalence rules in Table~\ref{tab:prop-equiv}-\ref{tab:pred-equiv}, logic inference rules in Table~\ref{tab:inference-law}, and common logical fallacies in Table~\ref{tab:common-fallacies}.

\begin{table*}[ht]
\small
\centering
\caption{Propositional logic equivalence laws.}
\label{tab:prop-equiv}
\begin{tabularx}{\textwidth}{@{}llX@{}}
\toprule
\bf Law & \bf  Logical Equivalence & \bf Example \\
\midrule
Idempotent laws & $P \land P \Leftrightarrow P$ & I am a teacher and I am a teacher $\Leftrightarrow$ I am a teacher.\\
                & $P \lor P \Leftrightarrow P$ & It's raining or it's raining $\Leftrightarrow$ it's raining.\\ \midrule
Commutative laws & $P \land Q \Leftrightarrow Q \land P$ & It is cold and it is winter $\Leftrightarrow$ It is winter and it is cold. \\
                 & $P \lor Q \Leftrightarrow Q \lor P$ & You can go to the party or you can study $\Leftrightarrow$ You can study or you can go to the party.\\ \midrule
Associative laws & $(P \land Q) \land R \Leftrightarrow P \land (Q \land R)$ & It is raining and it is cold, and also it is winter $\Leftrightarrow$ It is raining, and also, it is cold and it is winter.\\
                 & $(P \lor Q) \lor R \Leftrightarrow P \lor (Q \lor R)$ & Either I will go to the park or I will go to the library is true, or I will go to the cinema $\Leftrightarrow$ I will go to the park or either I will go to the library or I will go to the cinema is true.\\ \midrule
Distributive laws & $P \land (Q \lor R) \Leftrightarrow (P \land Q) \lor (P \land R)$ & It is raining and either I have an umbrella or I have a raincoat $\Leftrightarrow$ It is raining and I have an umbrella, or it is raining and I have a raincoat.\\
                  & $P \lor (Q \land R) \Leftrightarrow (P \lor Q) \land (P \lor R)$ & Either I will go to the park, or it is cloudy and it is cold $\Leftrightarrow$ Either I will go to the park or it is cloudy is true, and either I will go to the park or it is cold is true.\\\midrule
DeMorgan's laws & $\neg(P \land Q) \Leftrightarrow \neg P \lor \neg Q$ & It is not true that it's both cold and raining $\Leftrightarrow$ It's not cold or it's not raining.\\
                 & $\neg(P \lor Q) \Leftrightarrow \neg P \land \neg Q$ & It's not true that I will study or play $\Leftrightarrow$ I won't study and I won't play.\\\midrule
Complement laws & $\neg(\neg P) \Leftrightarrow P$ & It is not the case that it is not raining $\Leftrightarrow$ It is raining.\\\midrule
Conditional laws & $P \rightarrow Q \Leftrightarrow \neg P \lor Q$ & If it rains, then I'll stay at home $\Leftrightarrow$ It doesn't rain or I stay at home. \\
Bidirectional laws & $(P \leftrightarrow Q) \Leftrightarrow (P \land Q) \lor (\neg P \land \neg Q)$ & I'll go to the park if and only if it's sunny $\Leftrightarrow$ Either it's sunny and I go to the park, or it's not sunny and I don't go to the park.\\

\bottomrule
\end{tabularx}

\end{table*}

\begin{table*}[ht]
\small
\centering
\caption{Predicate logic quantifier laws.}
\label{tab:pred-equiv}


\begin{tabularx}{\textwidth}{@{}llX@{}}
\toprule
Law & Logical Equivalence & Example \\
\midrule
Quantifier Negation & $\neg \forall x P(x) \Leftrightarrow \exists x \neg P(x)$ & It is not the case that all birds can fly $\Leftrightarrow$ There exists a bird that cannot fly.\\
                            & $\neg \exists x P(x) \Leftrightarrow \forall x \neg P(x)$ & There is no human that can live forever $\Leftrightarrow$ All humans cannot live forever.\\\midrule
Quantifier Distribution & $\forall x (P(x) \land Q(x)) \Leftrightarrow \forall x P(x) \land \forall x Q(x)$ & Every student is smart and diligent $\Leftrightarrow$ Every student is smart, and every student is diligent.\\
                                & $\exists x (P(x) \lor Q(x)) \Leftrightarrow \exists x P(x) \lor \exists x Q(x)$ & There is a person who is either a doctor or a lawyer $\Leftrightarrow$ There is a person who is a doctor, or there is a person who is a lawyer.\\\midrule

Quantifier Movement & $\forall x (P \rightarrow Q(x)) \Leftrightarrow (P \rightarrow \forall x Q(x))$ & For every child, if it is raining then they are inside $\Leftrightarrow$ If it is raining, then every child is inside when the notion of raining doesn't depend on the specific child.\\
                    & $\exists x (P \land Q(x)) \Leftrightarrow (P \land \exists x Q(x))$ & There exists a student who is tall and a good basketball player $\Leftrightarrow$ There is a tall student and there exists a student who is a good basketball player when the notion of being tall doesn't depend on the specific student.\\
\bottomrule
\end{tabularx}

\end{table*}

\begin{table*}[ht]
\centering
\caption{Propositional and predicate logic inference rules.}
\label{tab:inference-law}
\small
\begin{tabularx}{\textwidth}{@{}llX@{}}
\toprule
Inference Rule & Logical Form & Example \\
\midrule
Universal Instantiation & $\forall x P(x) \vdash P(c)$ & All humans are mortal. Hence, Socrates is mortal. \\ 
Existential Generalization  & $P(c) \vdash \exists x P(x)$ & This apple is red. Hence, there exists a red apple. \\
Universal Generalization & $\{P(x)\} \vdash \forall y P(y)$ & Any particular human is mortal. Hence, all humans are mortal. \\
Modus Ponens & $\{P \to Q, P\} \vdash Q$ & If it rains, the street gets wet. It is raining. Hence, the street is wet. \\
Modus Tollens & $\{P \to Q, \neg Q\} \vdash \neg P$ & If it rains, the street gets wet. The street is not wet. Hence, it is not raining. \\
& $\{P \to \neg Q, Q\} \vdash \neg P$ & If it rains, the street does not get wet. The street is wet. Hence, it is not raining. \\
Transitivity  & $\{P \to Q, Q \to R\} \vdash P \to R$ & If I study, I will pass the test. If I pass the test, I will get a reward. Hence, if I study, I will get a reward. \\
Disjunctive Syllogism & $\{P \lor Q, \neg P\} \vdash Q$ & Either it's raining or it's snowing. It's not raining. Hence, it's snowing. \\
Addition & $\{P\} \vdash P \lor Q$ & It is raining. Hence, it is raining or snowing. \\
Simplification & $\{P \land Q\} \vdash P$ & It is raining and it is cold. Hence, it is raining. \\
& $\{P \land Q\} \vdash Q$ & It is raining and it is cold. Hence, it is cold. \\
Conjunction & $\{P, Q\} \vdash P \land Q$ & It is raining. It is cold. Hence, it is raining and it is cold. \\
Resolution & $\{P \lor Q, \neg P \lor R\} \vdash Q \lor R$ & Either it is raining or snowing. If it is not raining, then it is cloudy. Hence, either it is snowing or it is cloudy. \\
Disjunction Elimination & $\{P \to R, Q \to R, P \lor Q\} \vdash R$ & If it rains, I will stay home. If it snows, I will stay home. Either it will rain or snow. Hence, I will stay home. \\
Biconditional Introduction & $\{P \to Q, Q \to P\} \vdash P \leftrightarrow Q$ & If I study, I pass. If I pass, I studied. Hence, I study if and only if I pass. \\
Biconditional Elimination & $\{P \leftrightarrow Q\} \vdash P \to Q$ & I study if and only if I pass. Hence, if I study, I pass. \\
& $\{P \leftrightarrow Q, \neg P\} \vdash \neg Q$ & I study if and only if I pass. I didn’t study. Hence, I didn’t pass. \\
& $\{P \leftrightarrow Q, \neg Q\} \vdash \neg P$ & I study if and only if I pass. I didn’t pass. Hence, I didn’t study. \\

\bottomrule
\end{tabularx}
\end{table*}

\begin{table*}[ht]
\footnotesize
\centering
\caption{Common fallacies.}

\label{tab:common-fallacies}

\begin{tabularx}{\textwidth}{@{}llX@{}}
\toprule
Name & Logical Form & Example \\
\midrule
Affirming the Consequent & $p \to q, q \vdash p$ & If I study, I will pass the test. I passed the test. Therefore, I studied. \\
Denying the Antecedent & $p \to q, \neg p \vdash \neg q$ & If it rains, the street gets wet. It is not raining. Therefore, the street is not wet. \\
Affirming a Disjunct & $p \lor q, p \vdash \neg q$ & Either I will study or I will fail the test. I studied. Therefore, I will not fail the test. \\
Denying a Conjunct & $\neg (p \land q), \neg p \vdash q$ & I'm not both hungry and thirsty. I'm not hungry. Therefore, I'm thirsty. \\
Illicit Commutativity & $p \to q \vdash q \to p$ & If I am in Paris, then I am in France. Therefore, if I am in France, I am in Paris. \\
Existential Fallacy & $\forall x (P(x) \to Q(x)), \neg \exists x (P(x)) \vdash \neg \exists x (Q(x))$ & All birds can fly. No birds are present. Therefore, nothing can fly. \\
Illicit Major & $\forall x (P(x) \to Q(x)), \exists x (Q(x)) \vdash \exists x (P(x))$ & All humans are mortal. Something is mortal. Therefore, something is human. \\
Illicit Minor & $\forall x (P(x) \to Q(x)), \forall x (P(x) \to R(x)) \vdash \forall x (R(x) \to Q(x))$ & All men are mortal. All men are humans. Therefore, all humans are mortal. \\
Undistributed Middle & $\forall x (P(x) \to Q(x)), Q(a) \vdash P(a)$ & All dogs are animals. My cat is an animal. Therefore, my cat is a dog. \\
\bottomrule
\end{tabularx}

\end{table*}

\section{Extended Rules}
\label{appendix:extended-rules}
\subsection{Equivalent Extension}
The equivalent rule extension is based on the following fact: $$\{A \Leftrightarrow B, \forall x (A)\} \vdash \{\forall x (B)\}$$ (i.e., if A and B are equivalent, and for all x, A is true, then for all x, B is also true), and $$\{A \Leftrightarrow B, \exists x (A)\} \vdash \{\exists x (B)\}$$ (i.e., if A and B are equivalent, and there exist x such that A is true, then there exist x such that B is true). For example, the predicate version of the DeMorgan's law $$\neg(P \land Q) \Leftrightarrow \neg P \lor \neg Q$$ will become $$\forall x (\neg(P(x) \land Q(x))) \Leftrightarrow \forall x (\neg P(x) \lor \neg Q(x)),$$ and $$\exists x (\neg(P(x) \land Q(x))) \Leftrightarrow \exists x (\neg P(x) \lor \neg Q(x)).$$ In this example, the goal is to extend the propositional equivalence law to its predicate version by adding quantifiers. To achieve this goal, we first note that DeMorgan's law states that "P and Q cannot both be true" (e.g., Alice is happy and Bob is happy cannot both be true) is equivalent to "either not P or not Q" (e.g., either Alice is not happy or Bob is not happy). Since the two expressions are equivalent, we can add the same quantifier to both sides and the equivalence will still hold. Therefore, by adding a "for all" quantifier to both sides, we obtain "for all x, P(x) and Q(x) cannot both be true" (for all persons in the room, the person likes Charley and the person likes David cannot both be true) is equivalent to "for all x, either not P(x) or not Q(x)" (e.g., for all person in the room, either the person doesn't like Charley or the person doesn't like David). Before the extension, the law can only be applied to simple propositions (e.g., P = "Alice is happy", Q = "Bob is happy"), but after extension, the law can be applied to predicates with variables and quantifiers (e.g., P(x) = "x likes Charley", Q(x) = "x likes David") The same also applies to the "exist" quantifier.

\subsection{Inference Extension}
The inference rule extension is based on the following fact: $$\{A \land B \rightarrow C\} \vdash \{\forall x, (A) \land \forall x, (B) \rightarrow \forall x, (C)\},$$ (i.e., if A and B imply C, then for all x, A is true and for all x, B is true implies for all x, C is true) $$\{A \land B \rightarrow C\} \vdash \{\exists x, (A) \land \forall x, (B) \rightarrow \exists x, (C)\}.$$ (i.e., if A and B imply C, there exists x such that A is true and for all x, B is true implies there exists x such that C is true). Since all propositional inference rules are of the form $P \land Q \rightarrow C$, we can transform them into their predicate form $\forall x, P(x) \land \forall x, Q(x) \rightarrow \forall x, C(x)$ and $\exists x, P(x) \land \forall x, Q(x) \rightarrow \exists x, C(x)$ following similar procedure in the previous section.

\section{Natural Language Translation}
\label{appendix:translation}
\subsection{Algorithm}
Given an input: a logic clause of the form \texttt{[operator, Clause$_A$, Clause$_B$]}, where the clauses are also of the form \texttt{[operator, Clause$_A$, Clause$_B$]}, the algorithm will do the following:

\begin{enumerate}
    \item \textbf{Single Proposition Clause:} If the clause is just a single proposition, the algorithm finds this proposition's natural language form and returns it. The natural language form is obtained by combining vocabularies according to certain templates (e.g., subject + action).
    
    \item \textbf{Negation:} If the clause starts with a ``$\neg$'' operator, the algorithm then translates the rest of the clause based on a negation template, making sure to negate the statement.
    
    \item \textbf{Quantifiers:} For clauses that start with ``$\forall$'' (meaning for all items) or ``$\exists$'' (meaning there is at least one item), it translates these into natural language, adjusting the phrasing based on whether we're asserting something positively or negating it.
    
    \item \textbf{Logical Connectives:} If the clause combines propositions using logical operators like ``$\land$'', ``$\lor$'', ``$\rightarrow$'' (implies), or ``$\leftrightarrow$'' (if and only if), the function translates these into natural language phrases that express the relationship between the propositions.
\end{enumerate}

\subsection{Example}

Consider the expression: \texttt{[$\forall x$, $\rightarrow$, A($x$), B($x$)]}. Here's how the function would translate it:

\begin{enumerate}
    \item It sees the ``$\forall x$'' quantifier and adds ``For all $x$,'' to the sentence and continues to process the clause \texttt{[$\rightarrow$, A($x$), B($x$)]}.
    
    \item It sees the ``$\rightarrow$'' operator, which means ``if...then...''. It connects the two operands with the operator and obtains ``For all $x$, if A($x$), then B($x$)''. Then, it continues to process the clauses A($x$), B($x$).
    
    \item Since A($x$), B($x$) are single proposition clauses, the function looks up the vocabulary and synthesizes the natural language versions of the proposition. For example, A($x$) = ``$x$ drinks water'', B($x$) = ``$x$ is a cashier''.
    
    \item It constructs the sentence: ``For all $x$, if $x$ drinks water, then $x$ is a cashier''.
\end{enumerate}

\subsection{Vocabulary}
We list the vocabulary used in our experiment:
\subsection*{Subjects}
\begin{itemize}
    \item x, y, z, James, Mary, Robert, Patricia, John, Jennifer, Michael, Linda, William, Elisabeth, David, Barbara, Richard, Susan, Joseph, Jessica, Thomas, Sarah, Charles, Karen, Alice, Benjamin, Daniel, Emily, George, Helen, Ian, Julie.
\end{itemize}

\subsection*{Predicates}
\begin{itemize}
    \item a cashier, a janitor, a bartender, a server, an office clerk, a mechanic, a carpenter, an electrician, a nurse, a doctor, a police officer, a taxi driver, a soldier, a politician, a lawyer, a scientist, an astronaut, a poet, an artist, a sailor, a writer, a musician, poor, rich, happy, sad, fast, curious, excited, bored, tired, joyful, intelligent, skilled, efficient, meticulous, creative.
\end{itemize}

\subsection*{Actions}
\begin{itemize}
    \item make tea, makes tea, making tea, drink water, drinks water, drinking water, read a book, reads a book, reading a book, play tennis, plays tennis, playing tennis, play squash, plays squash, playing squash, play a game, plays a game, playing a game, go running, goes running, running, work, works, working, sleep, sleeps, sleeping, cook, cooks, cooking, listen to a song, listens to a song, listening to a song, write a letter, writes a letter, writing a letter, drive a car, drives a car, driving a car, climb a mountain, climbs a mountain, climbing a mountain, take a plane, takes a plane, taking a plane, paint a picture, paints a picture, painting a picture.
\end{itemize}

\subsection*{Impersonal Candidates}
\begin{itemize}
    \item snowing, snows, doesn't snow, snow, raining, rains, doesn't rain, rain, sunny, is sunny, is not sunny, be sunny, cloudy, is cloudy, is not cloudy, be cloudy, windy, is windy, is not windy, be windy, cold, is cold, is not cold, be cold, late, is late, is not late, be late, overcast, is overcast, is not overcast, be overcast, foggy, is foggy, is not foggy, be foggy, humid, is humid, is not humid, be humid.
\end{itemize}

\section{Prompting LLMs}
\label{appendix:prompting}
For all GPT models, we set the system prompt of to blank. 

\subsection{Zero-Shot Example}

\texttt{Consider the following premises: The claim that John is a poet and the claim that it is cloudy cannot both be true. Can we infer the following from them? Answer yes or no: Jessica is not listening to a song.}

\subsection{Zero-Shot CoT}
For Zero-Shot CoT, we add \texttt{"Please think step-by-step and answer the following question."} To zero-shot queries.

\subsection{ICL Example}
\texttt{Q: Consider the following premises: If it snows, then Joseph is a politician. Can we infer the following from them? Answer yes or no: It is snowing if and only if Joseph is a politician. \\\\ A: Let A be the claim that "it snows", B be the claim that "Joseph is a politician", then the premises are "if A then B, if B then A", which is equivalent to "A if and only if B" by the biconditional introduction rule. Therefore, we can infer that It is snowing if and only if Joseph is a politician. The answer is ==yes==.\\\\Q: Consider the following premises: It is late and it is windy. Can we infer the following from them? Answer yes or no: It is windy.\\\\A: Let A be the claim that "it is late", B be the claim that "it is windy", then the premises are "A and B". By the simplification rule, we can infer B. Therefore, we can infer that it is windy. The answer is ==yes==.\\\\Q: Consider the following premises: Jessica is not running. Jessica is running or it is raining. Can we infer the following from them? Answer yes or no: Sarah is not happy.\\\\A: Let A be the claim that "Jessica is running", B be the claim that "it is raining", C be the claim that "Sarah is happy", then the premises are "not A, A or B". We cannot infer C from the premises. The answer is ==no==.\\\\Q: Consider the following premises: It is not raining. It is raining or it is late. Can we infer the following from them? Answer yes or no: It is not late.\\\\A: Let A be the claim that "it is raining", B be the claim that "it is late", then the premises are "not A, A or B". We can infer "B", which is "it is late". Therefore, we cannot infer "it is not late". The answer is ==no==.\\\\Q: Consider the following premises: For all x, x will write a letter, and x will climb a mountain and x is a musician. Can we infer the following from them? Answer yes or no: For all x, x will write a letter and x will climb a mountain, and x is a musician.}

\section{Accessing LLMs}
\label{appendix:access-LLMs}
For commercial models, the specific models we accessed are \texttt{gpt-4o-2024-05-13} for GPT-4o, \texttt{gpt-4-turbo-2024-04-09} for GPT-4, \texttt{gpt-3.5-turbo-0125} for ChatGPT, \texttt{gemini-1.5-flash-latest} (accessed June, 2024) for Gemini-1.5. All accesses are made via their official APIs~\footnote{\url{https://platform.openai.com/docs/} and \url{https://ai.google.dev/gemini-api}}. For the open-source models, we use their respective Hugging Face~\footnote{\url{https://huggingface.co/}} repository, i.e., \texttt{meta-llama/Meta-Llama-3-70B} for Llama-3 and \texttt{mistralai/Mixtral-8x7B-v0.1} for Mixtral-8x7B. 

For model parameters, we set the \textit{temperature} to 0.0 and \textit{max\_tokens} to 500 for all models. We keep other parameters to the models' respective default values.

\section{Sampled Data Statistics}
\label{appendix:stat}
We analyzed various statistics of the sampled dataset, classified into different categories as presented in Table \ref{tab:distribution}. The table summarizes the distribution of logic types, categorizes the rule types used in our analysis, and details the types of problems.

\begin{table}[h]
    \centering
    \caption{Distribution of logic types, rule categories, and problem types.}
    \label{tab:distribution}
    \begin{tabular}{ll}
        \toprule
        \midrule
        \textbf{Logic Type} & \textbf{Count} \\
        Predicate           & 146          \\
        Propositional       & 62           \\
        \midrule
        \textbf{Rule Category} & \textbf{Count} \\
        Inference              & 108          \\
        Equivalent             & 81           \\
        Fallacy                & 19           \\
        \midrule
        \textbf{Problem Type} & \textbf{Count} \\
        Inference             & 82            \\
        Unrelated             & 63            \\
        Contradiction         & 63            \\
        \midrule
        Total & 208\\
        \bottomrule
    \end{tabular}
\end{table}

\section{Accuracy Versus Inference Length}
\label{appendix:lengths}
To assess the impact of inference length, we generated test cases of varying
lengths (i.e., ranging from 1 to 9) using randomly selected rules. For each length, we generated 100 test cases. Table~\ref{tab:length} shows the performance of LLMs in these test cases. Generally, LLMs perform gradually worse as the inference length increases, indicating the increased complexity introduced by longer inference chains.

\begin{table}[ht]
    \centering
    \caption{LLMs' performance versus inference length.}
    \label{tab:length}
\begin{tabular}{lcccccc}
        \toprule
        Len & 1 & 3 & 5 & 7 & 9 \\
        \midrule
        GPT4    & 0.99 & 0.88 & 0.81 & 0.86 & 0.76 \\
        GPT4o   & 0.94 & 0.81 & 0.79 & 0.72 & 0.63 \\
        Gemini  & 0.93 & 0.80 & 0.73 & 0.74 & 0.64 \\        
        Llama3  & 0.91 & 0.78 & 0.74 & 0.81 & 0.67 \\
        Mixtral & 0.79 & 0.67 & 0.64 & 0.66 & 0.62 \\
        ChatGPT & 0.80 & 0.77 & 0.71 & 0.61 & 0.44 \\
        \bottomrule
    \end{tabular}

\end{table}

\section{Complete Break-Down Result for All LLMs} 
\label{appendix:breakdown-list}
In this section, we list the complete results of all LLMs on the set of 208 atomic rules in Table~\ref{tab:breakdown-gpt4} through Table~\ref{tab:breakdown-chatgpt}. The results are sorted in ascending order according to the zero-shot performance of the models.
\newpage

\centering
\onecolumn
\scriptsize
\begin{longtable}{llllllll}
\caption{Break-down of the accuracy of GPT-4 on all rules (sorted by zero-shot accuracy).} \label{tab:breakdown-gpt4} \\
\toprule
Logic & Rule category & Rule & Problem & Zero shot & Zero shot cot & Random icl & Weak \\
\midrule
\endfirsthead
\caption[]{Break-down of the accuracy of GPT-4 on all rules (sorted by zero-shot accuracy).} \\
\toprule
Logic & Rule category & Rule & Problem & Zero shot & Zero shot cot & Random icl & Weak \\
\midrule
\endhead
\midrule
\multicolumn{8}{r}{Continued on next page} \\
\midrule
\endfoot
\bottomrule
\endlastfoot
Predicate & Inference & Universal generalization & Inference & 0.52 & 0.52 & 0.52 & 0.52 \\
Predicate & Inference & Existential resolution & Unrelated & 0.60 & 0.76 & 1.00 & 1.00 \\
Predicate & Inference & Universal resolution & Inference & 0.60 & 0.36 & 0.08 & 1.00 \\
Predicate & Equivalent & Law of quantifier movement & Inference & 0.68 & 0.60 & 0.60 & 0.72 \\
Predicate & Inference & Existential resolution & Inference & 0.72 & 0.52 & 0.96 & 0.96 \\
Predicate & Inference & Existential biconditional introduction & Inference & 0.76 & 0.44 & 0.96 & 1.00 \\
Predicate & Inference & Existential biconditional introduction & Contradiction & 0.80 & 0.80 & 1.00 & 1.00 \\
Predicate & Equivalent & Existential conditional laws & Inference & 0.80 & 0.80 & 1.00 & 1.00 \\
\end{longtable}

\scriptsize
\begin{longtable}{llllllll}
\caption{Break-down of the accuracy of GPT-4o on all rules (sorted by zero-shot accuracy).} \label{tab:breakdown-gpt4o} \\
\toprule
Logic & Rule category & Rule & Problem & Zero shot & Zero shot cot & Random icl & Weak \\
\midrule
\endfirsthead
\caption[]{Break-down of the accuracy of GPT-4o on all rules (sorted by zero-shot accuracy).} \\
\toprule
Logic & Rule category & Rule & Problem & Zero shot & Zero shot cot & Random icl & Weak \\
\midrule
\endhead
\midrule
\multicolumn{8}{r}{Continued on next page} \\
\midrule
\endfoot
\bottomrule
\endlastfoot
Propositional & Inference & Resolution & Inference & 0.04 & 0.64 & 1.00 & 1.00 \\
Predicate & Inference & Universal resolution & Inference & 0.08 & 0.00 & 0.16 & 0.88 \\
Predicate & Inference & Existential biconditional introduction & Unrelated & 0.20 & 0.96 & 1.00 & 0.56 \\
Propositional & Inference & Biconditional introduction & Unrelated & 0.40 & 1.00 & 1.00 & 1.00 \\
Propositional & Fallacy & Denying a conjunct & Inference & 0.40 & 1.00 & 1.00 & 1.00 \\
Predicate & Fallacy & Existential denying the antecedent & Inference & 0.40 & 1.00 & 0.96 & 0.88 \\
Predicate & Inference & Universal biconditional introduction & Unrelated & 0.40 & 1.00 & 1.00 & 1.00 \\
Predicate & Inference & Existential resolution & Inference & 0.52 & 0.00 & 0.00 & 0.56 \\
\end{longtable}

\scriptsize
\begin{longtable}{llllllll}
\caption{Break-down of the accuracy of Gemini-1.5 on all rules (sorted by zero-shot accuracy).}\label{tab:breakdown-gemini}  \\
\toprule
Logic & Rule category & Rule & Problem & Zero shot & Zero shot cot & Random icl & Weak \\
\midrule
\endfirsthead
\caption[]{Break-down of the accuracy of Gemini-1.5 on all rules (sorted by zero-shot accuracy).} \\
\toprule
Logic & Rule category & Rule & Problem & Zero shot & Zero shot cot & Random icl & Weak \\
\midrule
\endhead
\midrule
\multicolumn{8}{r}{Continued on next page} \\
\midrule
\endfoot
\bottomrule
\endlastfoot
Predicate & Fallacy & Existential denying the antecedent & Inference & 0.00 & 0.60 & 0.48 & 0.84 \\
Predicate & Inference & Existential biconditional introduction & Unrelated & 0.08 & 0.28 & 0.12 & 0.32 \\
Predicate & Inference & Existential biconditional introduction & Contradiction & 0.16 & 0.52 & 0.96 & 0.88 \\
Predicate & Inference & Existential resolution & Unrelated & 0.20 & 0.40 & 0.44 & 0.80 \\
Propositional & Fallacy & Denying the antecedent & Inference & 0.32 & 0.80 & 0.92 & 0.72 \\
Predicate & Fallacy & Existential denying a conjunct & Inference & 0.32 & 0.40 & 0.64 & 0.88 \\
Predicate & Inference & Universal instantiation & Inference & 0.32 & 1.00 & 0.64 & 0.64 \\
Predicate & Inference & Universal disjunctive syllogism & Contradiction & 0.44 & 1.00 & 1.00 & 1.00 \\
\end{longtable}

\scriptsize
\begin{longtable}{llllllll}
\caption{Break-down of the accuracy of Llama3 on all rules (sorted by zero-shot accuracy).} \label{tab:breakdown-llama3}  \\
\toprule
Logic & Rule category & Rule & Problem & Zero shot & Zero shot cot & Random icl & Weak \\
\midrule
\endfirsthead
\caption[]{Break-down of the accuracy of Llama3 on all rules (sorted by zero-shot accuracy).} \\
\toprule
Logic & Rule category & Rule & Problem & Zero shot & Zero shot cot & Random icl & Weak \\
\midrule
\endhead
\midrule
\multicolumn{8}{r}{Continued on next page} \\
\midrule
\endfoot
\bottomrule
\endlastfoot
Predicate & Fallacy & Existential denying the antecedent & Inference & 0.00 & 0.72 & 0.44 & 0.40 \\
Predicate & Inference & Existential resolution & Unrelated & 0.04 & 0.28 & 0.16 & 0.12 \\
Predicate & Inference & Existential biconditional introduction & Unrelated & 0.08 & 0.64 & 0.20 & 0.04 \\
Predicate & Fallacy & Universal denying the antecedent & Inference & 0.12 & 0.76 & 0.76 & 0.52 \\
Predicate & Fallacy & Existential denying a conjunct & Inference & 0.16 & 0.16 & 0.60 & 0.48 \\
Predicate & Fallacy & Universal denying a conjunct & Inference & 0.16 & 0.76 & 0.52 & 0.12 \\
Propositional & Fallacy & Denying the antecedent & Inference & 0.28 & 0.88 & 0.52 & 0.16 \\
Predicate & Fallacy & Existential fallacy & Inference & 0.28 & 0.80 & 0.28 & 0.28 \\
\end{longtable}

\scriptsize
\begin{longtable}{llllllll}
\caption{Break-down of the accuracy of Mixtral on all rules (sorted by zero-shot accuracy).} \label{tab:breakdown-mixtral}  \\
\toprule
Logic & Rule category & Rule & Problem & Zero shot & Zero shot cot & Random icl & Weak \\
\midrule
\endfirsthead
\caption[]{Break-down of the accuracy of Mixtral on all rules (sorted by zero-shot accuracy).} \\
\toprule
Logic & Rule category & Rule & Problem & Zero shot & Zero shot cot & Random icl & Weak \\
\midrule
\endhead
\midrule
\multicolumn{8}{r}{Continued on next page} \\
\midrule
\endfoot
\bottomrule
\endlastfoot
Predicate & Inference & Universal disjunction elimination & Inference & 0.04 & 0.16 & 0.32 & 0.32 \\
Predicate & Inference & Universal modus ponens & Inference & 0.04 & 0.12 & 0.08 & 0.16 \\
Predicate & Equivalent & Existential conditional laws & Inference & 0.08 & 0.20 & 0.00 & 0.00 \\
Predicate & Inference & Universal modus tollens & Inference & 0.08 & 0.16 & 0.12 & 0.20 \\
Predicate & Equivalent & Universal conditional laws & Inference & 0.20 & 0.16 & 0.20 & 0.24 \\
Predicate & Inference & Universal addition & Inference & 0.24 & 0.08 & 0.04 & 0.00 \\
Predicate & Inference & Universal disjunctive syllogism & Inference & 0.28 & 0.20 & 0.20 & 0.32 \\
Predicate & Inference & Existential addition & Inference & 0.32 & 0.28 & 0.04 & 0.00 \\
\end{longtable}

\scriptsize
\begin{longtable}{llllllll}
\caption{Break-down of the accuracy of ChatGPT on all rules (sorted by zero-shot accuracy).} \label{tab:breakdown-chatgpt} \\
\toprule
Logic & Rule category & Rule & Problem & Zero shot & Zero shot cot & Random icl & Weak \\
\midrule
\endfirsthead
\caption[]{Break-down of the accuracy of ChatGPT on all rules (sorted by zero-shot accuracy).} \\
\toprule
Logic & Rule category & Rule & Problem & Zero shot & Zero shot cot & Random icl & Weak \\
\midrule
\endhead
\midrule
\multicolumn{8}{r}{Continued on next page} \\
\midrule
\endfoot
\bottomrule
\endlastfoot
Predicate & Inference & Existential resolution & Unrelated & 0.00 & 0.40 & 0.72 & 0.52 \\
Predicate & Equivalent & Universal de morgan's laws & Contradiction & 0.04 & 0.40 & 0.24 & 0.12 \\
Predicate & Equivalent & Existential de morgan's laws & Contradiction & 0.08 & 0.96 & 0.64 & 1.00 \\
Predicate & Inference & Existential biconditional introduction & Unrelated & 0.08 & 0.52 & 0.40 & 0.52 \\
Predicate & Inference & Universal instantiation & Inference & 0.08 & 0.44 & 0.04 & 0.12 \\
Predicate & Fallacy & Existential denying the antecedent & Inference & 0.12 & 0.88 & 0.76 & 0.68 \\
Predicate & Inference & Universal disjunctive syllogism & Contradiction & 0.12 & 0.60 & 0.20 & 0.04 \\
Predicate & Inference & Universal transitivity & Contradiction & 0.12 & 0.32 & 1.00 & 0.68 \\
\end{longtable}

\end{document}